\documentclass[twocolumn]{aastex62}

\usepackage{easyReview}
\usepackage{fontawesome}
\usepackage{hyperref}
\usepackage{amsmath}
\submitjournal{ApJL}
\accepted{September 29, 2019}

\shorttitle{Required Precision of Exoplanet Masses}
\shortauthors{Batalha et al.}

\begin{document}

\title{The Precision of Mass Measurements Required for Robust Atmospheric Characterization \\ of Transiting Exoplanets}

\correspondingauthor{Natasha E. Batalha}
\email{nbatalha@ucsc.edu}

\author{Natasha E. Batalha}
\affil{University of California, Santa Cruz, CA}

\author{Taylor Lewis}
\affil{University of California, Santa Cruz, CA}

\author{Jonathan J. Fortney}
\affiliation{University of California, Santa Cruz, CA}

\author{Natalie M. Batalha}
\affiliation{University of California, Santa Cruz, CA}

\author{Eliza Kempton}
\affil{Universty of Maryland, College Park, MD}

\author{Nikole K. Lewis}
\affil{Cornell University, Ithaca, NY}

\author{Michael R. Line}
\affil{Arizona State University, Tempe, AZ}



\begin{abstract}
Two of TESS's major science goals are to measure masses for 50 planets smaller than 4 Earth radii and to discover high-quality targets for atmospheric characterization efforts. It is important that these two goals are linked by quantifying what precision of mass constraint is required to yield robust atmospheric properties of planets. Here, we address this by conducting retrievals on simulated JWST transmission spectra under various assumptions for the degree of uncertainty in the planet's mass for a representative population of seven planets ranging from terrestrials to warm Neptunes to hot Jupiters.  Only for the cloud-free, low metallicity gas giants are we able to infer exoplanet mass from transmission spectroscopy alone, to $\sim$10\% accuracy. For low metallicity cases ($<$ 4$\times$ Solar) we are able to accurately constrain atmospheric properties without prior knowledge of the planet's mass.  For all other cases (including terrestrial-like planets), atmospheric properties can only be inferred with a mass precision of better than $\pm$50\%. At this level, though, the widths of the posterior distributions of the atmospheric properties are dominated by the uncertainties in mass. With a precision of $\pm$20\%, the widths of the posterior distributions are dominated by the spectroscopic data quality. Therefore, as a rule-of-thumb, we recommend: a $\pm$50\% mass precision for initial atmospheric characterization and a $\pm$20\% mass precision for more detailed atmospheric analyses.
\end{abstract}



\section{Introduction} \label{sec:intro}

NASA's Transiting Exoplanet Survey Satellite (\textit{TESS}) has already identified fruitful targets for atmospheric characterization with \textit{Hubble} as well as the James Webb Space Telescope (\textit{JWST}). In addition to providing targets for atmospheric characterization, one of TESS's Level One science requirement is to measure masses for 50 transiting planets smaller than 4 Earth radii. Ground-based follow-up campaigns are well-underway, and the question arises: how precisely do observes need to determine a planet's mass in order to robustly characterize the atmosphere?  Setting quantitative guidelines on mass precision and accuracy via simulations will help us avoid wasting precious time on ground-based facilities. Such guidelines will be relevant to future missions as well such as \textit{PLATO}. 

A full suite of ground based radial velocity facilities are already working together to optimize science yield for these detection missions. Yet, obtaining masses, especially for low-mass planets, is technically challenging and time intensive for both transit timing variations and radial velocity techniques. 

Generally, masses are obtained before atmospheric reconnaissance. However, a tension exists between the desire to study the atmospheres of recently discovered exoplanets and the goal of carefully vetting and selecting the optimal set of planets for such studies. Currently, there is no clear standard as to how precise planet masses should/need be before in-depth atmospheric studies. 

\begin{figure*}[t]
\centering
\includegraphics[width=\textwidth]{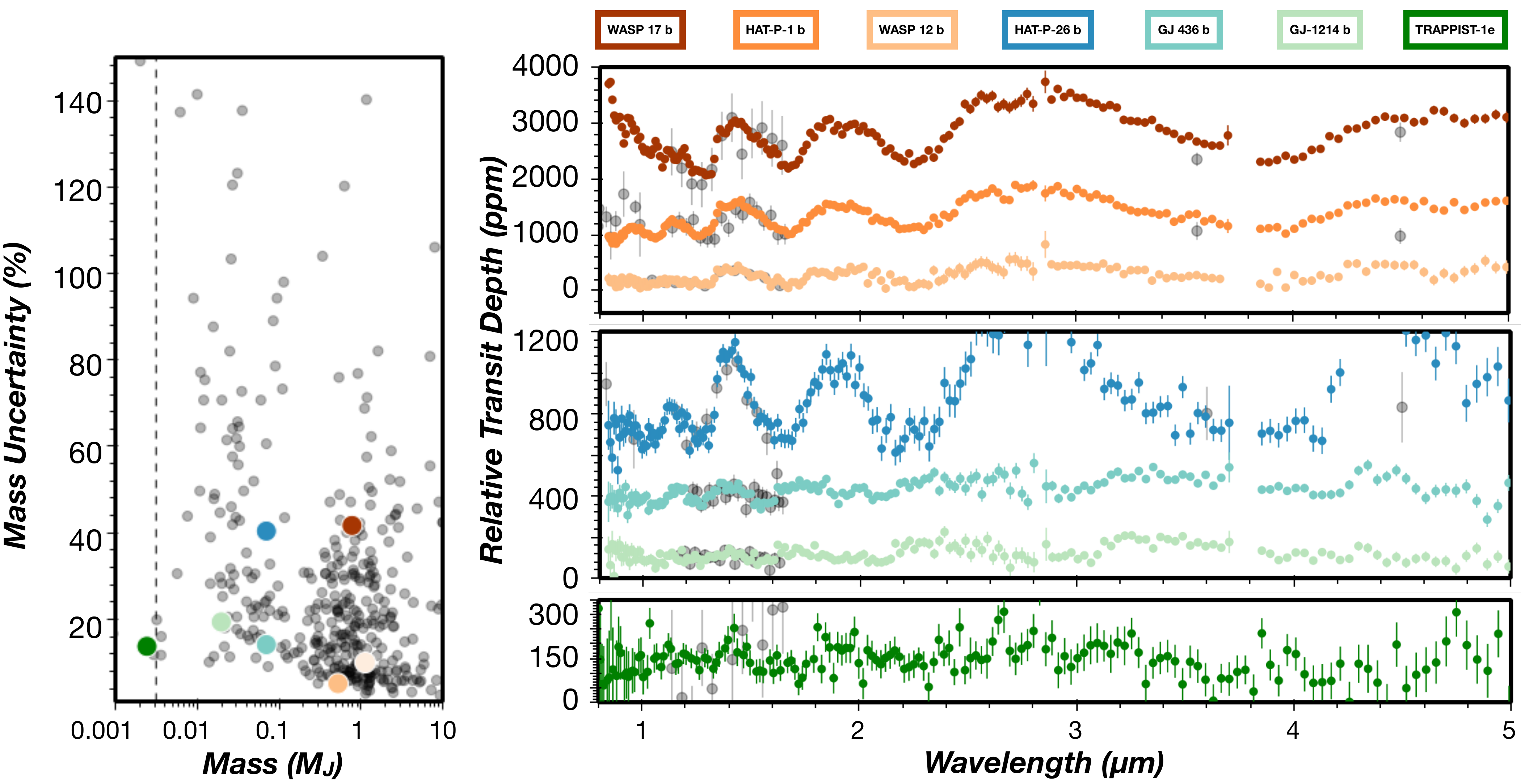}
\caption{The left hand plot shows the precision of measured masses for the sample of known exoplanets as a function of their best-fit mass.  The vertical dashed line indicates 1 Earth mass.  The right hand plot shows published \textit{Hubble} data in grey and simulated \textit{JWST} data (from this work) in color for seven representative planets that we study in detail in this paper.  The same seven planets' masses and associated uncertainties are also plotted in the lefthand panel using the same color scheme.  \textbf{Main Point:} This group of objects covers a diverse set of planet type, ranging from terrestrial-sized, to sub-Neptune, and hot Jupiters.}
\label{fig:sample} 
\end{figure*}

Figure \ref{fig:sample} shows the exoplanets for which mass measurements have been obtained, along with the uncertainty of each mass measurement. The majority of planets with mass measurements with worse than 10\% uncertainty are greater than 0.1~M$_\mathrm{J}$. However, even large planet masses ($>$0.1~M$_\mathrm{J}$) are sometimes are only precise to $\sim$50\% or greater.

Previous work using a retrieval framework suggested that the mass of a transiting exoplanet could be directly inferred from its high signal-to-noise (SNR), \textit{JWST}-like, transmission spectrum alone \citep[e.g.][]{dewit2013constraining}. The method leverages the effect of the planet's surface gravity on the atmospheric scale height, which in turn influences the transmission spectrum. Later, forward models of smaller R=1.5R~$_\oplus$-sized planets from \citet{batalha2017challenges} demonstrated that degeneracies exist between transmission spectra of planets with different masses and compositions, making it difficult to unambiguously determine the planet’s mass and composition in many cases. 

While the specific role of uncertain mass measurements has not been well-explored, there has been a rich literature on the role of degeneracies in interpreting transmission spectroscopy, ranging from semi-analytical studies \citep[e.g.][]{des2008rayleigh,heng2017theory} to retrieval analyses \citep[e.g.][]{benneke2012atmospheric,benneke2013how, line2016clouds,bet2016degen,bet2017degen,macdonald2017hd209, welbanks2019degen}. For example, \citet{benneke2013how} investigated degeneracies between mean molecular weight and cloud top pressure, and \citet{heng2017theory} \& \citet{welbanks2019degen} focused on the degeneracy between radius, reference pressure, and composition. 

Despite these comprehensive analyses, little work has been done to study the effect of measured mass uncertainty on retrieved atmospheric parameters in a systematic way. We aim to answer three specific questions:
\begin{enumerate}
    \item To what level of precision (if any at all) do mass measurements need to be made for robust retrieval of atmospheric properties? 
    \item Does this level of precision vary by planet size or by atmospheric metallicity? Does it vary according to the precision attained on the transmission spectrum?
    \item Given poor to no initial mass measurement (i.e. upper limit), could our knowledge of mass be improved from transmission spectroscopy alone? 
\end{enumerate}

\begin{table*}[t]
    \begin{center}
    \begin{tabular}{c|c|c|c|c|c|c|c|c|c|c} \hline
        Planet & \textit{Hubble} Data\footnote{\textit{Hubble} data and characterization paper used to callibrate forward models and JWST simulations shown in Figure 1} & $M_{Jup}$ & $R_{Jup}$ & T$_{iso}$[K]\footnote{Isothermal temperature used for forward models based on characterization paper} & M/H\footnote{Atmospheric metallicity used for forward models based on characterization paper, when possible} & log$\kappa_\mathrm{cld}$ \footnote{Grey cloud opacity with an abundance-weighted cross section in units of log m$^2$/particle} & K mag & $R_\odot$ & $n_{occ}$\footnote{Number of transits per observation mode used for \textit{JWST} simulations} & $\sigma_{R=100}$ \footnote{R=100 precision, in units of ppm, at 1.4$\mu$m of JWST simulation, including all transit observations} \\ \hline \hline
        WASP 17b & \citet{mandell2013wasp17} & 0.78 & 1.87 & 1755 &  1$\times$Solar & -31 & 10.2 & 1.49 & 1 & 35 \\  
        HAT-P-1b & \citet{wakeford2013hatp1} & 0.525 & 1.319 & 1322 &  1$\times$Solar & -30.5 & 8.9 & 1.17 & 1 & 30 \\
        WASP 12b & \citet{kreidberg2015wasp12} & 1.31 & 1.82 & 1400 &  1$\times$Solar & -29.5 & 10.2 & 1.17 & 1 & 45 \\       
        HAT-P-26b & \citet{wakeford2017hatp26} & 0.06 & 0.55 & 990 & 4.15$\times$Solar & -29 & 9.6 & 0.87 & 1 & 40 \\ 
        GJ 436b & \citet{knutson2014gj436} & 0.07 & 0.37 & 686 & 400$\times$Solar & -28 & 6.1 & 0.46 & 2 & 20 \\ 
        GJ 1214b & \citet{kreidberg2014gj1214} & 0.0197 & 0.254 & 547 & 1000$\times$Solar & -27.5 & 8.8 & 0.22 & 3 & 20 \\ TRAPPIST-1e & \citet{dewit2018trappy} & 0.002 & 0.081 & 250 & 90\% H$_2$O, 10\% H$_2$ & -26 & 10.3 & 0.12 & 5 & 35 \\ 
        \hline
    \end{tabular}
    \end{center}
    \caption{Case Studies}\label{tab:planets}
\end{table*}

\section{Methods} \label{sec:methods}
The sample of planets explored in this work (shown in Figure \ref{fig:sample}), was chosen to cover a large parameter space in stellar properties (stellar type and magnitude), planet properties (planetary radius and mass), and known atmospheric properties from \textit{Hubble} spectroscopy (temperature, metallicity, degree of cloud and scattering extinction). Table 1 shows the full list of planets with the most relevant planet and star properties, \textit{JWST} observing strategies, and atmospheric properties used for forward modeling. Our population consists of three hot Jupiters (WASP-17b, HAT-P-1b, WASP-12b), three Neptunes/sub-Neptunes (HAT-P-26 b, GJ 436b, GJ 1214b) and one temperate rocky planet (TRAPPIST-1e). 

For each of the seven cases we create transmission spectra that are consistent with existing \textit{Hubble} WFC3 G141 spectroscopy from $1.1-1.7$~$\mu$m (citations in Table 1, data in grey in Figure 1). Then we use those modeled transmission spectra to create full \textit{JWST} synthetic observations, spanning 0.7-5 $\mu$m. These synthetic observations are then used within a retrieval framework to determine how well atmospheric properties can be inferred. We emphasize the goal of this work is not to conduct a uniform retrieval analysis of all these systems. We only strive to be consistent with existing analyses from \textit{Hubble} WFC3 observations (demonstrated in Figure 1).

\subsection{Modeling Transmission Spectra}
Our model \citep[\texttt{CHIMERA}][]{line2012info,line2013systematic,line2014systematic} takes in as input, planet and stellar properties, correlated-K opacities, a pressure-temperature profile, chemical abundances, and cloud properties. 

Our pressure-temperature (PT) profiles are chosen to be consistent with previous retrieval studies. We assume isothermal profiles ($T_{iso}$ in Table 1) because our aim is only to study the effect of mass uncertainty on retrieved atmospheric profiles and our focus is only on transmission spectroscopy, which is less sensitive to the full PT profile. Even though isothermal profiles have been shown to bias retrieved molecular abundances \citep{rocchetto2016bias}, this assumption will not impact our results. To ensure this, we check our results by using the 5-parameter PT profile parameterization from \citet{guillot2010pt}.  

We assume that clouds are not wavelength-dependent across the wavelengths explored here (1-5 $\mu$m). We model this grey cloud opacity with an abundance-weighted cross section ($\kappa_\mathrm{cld}$ in Table~1, log m$^2$/particle). This simple methodology has been widely used to reproduce the cloud behavior of hot Jupiters \citep[e.g.][]{sing2016hjs,fisher201838planets}, and within the \texttt{CHIMERA} framework \citep{schlawin2018clear}.

Initial chemical profiles for all the planet cases besides TRAPPIST-1e are computed using the chemical equilibrium model grid\footnote{\href{https://github.com/ExoCTK/chimera/blob/master/TEMPLATE\_TRANSMISSION/chem_full.pic}{ExoCTK CHIMERA Chemistry grid}} described in detail in \citet{schlawin2018clear}. Briefly, the chemical equilibrium grid was computed using \texttt{CEA} \citep{mcbride1996computer} for a range of metallicities ($0.03$-1000$\times$Solar), carbon-to-oxygen ratios (C/O $=$ 0.1-2), temperatures (300-3000 K), and pressures (250 bars - 0.1 $\mu$bars). 

For all chemical equillibrium cases we assume solar C/O=0.55. The metallicities were chosen to remain consistent with retrieved results from the observation studies shown in Table~1, when possible. For GJ 436b, and GJ 1214b, whose \textit{Hubble} WFC3 observation present no spectral features, we chose medium (400$\times$Solar) and high (1000$\times$Solar) metallicities, respectively, in order to: 1) remain consistent with the flat behavior of the \textit{Hubble} WFC3 observations and 2) explore the effect of a range of mean molecular weights, similar to the methodology of \citet{Greene2016characterizing} and \citet{schlawin2018clear}. The grid produces abundances for various species including C$_2$H$_2$, CH$_4$, CO, CO$_2$, H$_2$O, H$_2$S, HCN, Na, K, VO, TiO. 

We account for the opacities of all of these, including H$_2$-H$_2$, H$_2$-He collision induced absorption. Using the opacities originally described in \citet{freedman2008opacities,freedman2014opacities} and \citet{lupu2014earth}, we use the resort-rebin method described in \citet{amundsen2017cork} to compute correlated-K tables at R=100.

Lastly, for TRAPPIST-1e, which has no detectable spectral features in the \textit{Hubble} WFC3 bandpass, we choose a mixture of H$_2$O and H$_2$ (90\%/10\%) similar to \citet{batalha2017challenges} and \citet{Greene2016characterizing}. Given the low mean molecular weight of H$_2$O as compared to a more Earth-like or a Venus-like atmosphere, the absolute magnitude with which we retrieve the molecular abundance for the TRAPPIST-1e case will likely be optimistic compared to previous retrieval analysis of terrestrial atmospheres \citep{barstow2016hab, benneke2012atmospheric,kt2018detect}.

\subsection{\textit{JWST} Simulations \& Retrieval Framework}
We use \texttt{PandExo} \citep{batalha2017pandexo} to simulate ``typical'' \textit{JWST} observations for each case. A precision that would exceed the noise floor, or that would inhibit basic atmospheric characterization would not allow us to determine the effect of mass precision on retrieved parameters. We ensure our results are insensitive to the number of transits by spot-checking cases with $\pm$2 transits. 

All targets except TRAPPIST-1e are simulated with both NIRISS SOSS and NIRSpec G395H to maximize spectral information content \citep{batalha2017ic}. For both SOSS and G395H we choose the duty cycle based on an 80\% full-well saturation level. The R=100 precision, $\sigma_{R=100}$ at 1~$\mu$m is shown in Table 1. We give all hot Jupiters a single transit per visit. GJ 4346b and GJ 1214b are given 2 and 3 visits per observing mode, respectively, in order achieve the plausible \textit{JWST} noise floor \citep[20 ppm][]{Greene2016characterizing} and the precision of the current WFC3 observations \citep{kreidberg2014gj1214, knutson2014gj436}. Note that the noise floor was only used to guide number of transits, not in the simulations themselves. Lastly, for TRAPPIST-1e, whose host star is relatively dim, we use 5 transits of NIRSpec Prism yielding a precision of $\sim$35 ppm across 1-5 $\mu$m. 

From the simulated \textit{JWST} observations we use the open-source Nested Sampler, \texttt{dynesty} \citep{dynesty} to conduct retrievals. For all cases we retrieve isothermal temperature, metallicity, C/O ratio, reference radius, cloud cross section, haze power law and haze cross section, and mass in some cases. 

The reference radius, $xR_p$, is defined as a scaling factor to the reported radius from transit observation ($R_p$), which we arbitrarily define at 10 bars. The prior on this scaling factor is broad and uniform from 0.5-1.5. Altitude-dependent gravity is then computed in the model using the mass, and the scaled reference radius multiplied by the reported radius from transit studies.

For the highest metallicity cases (TRAPPIST-1e and GJ 1214b) we directly retrieve abundances as opposed to using the chemical equilibrium grid, which stops at 1000$\times$~Solar. For all these properties we use broad uniform priors. 

\begin{figure*}
\centering
\includegraphics[width=0.8\textwidth]{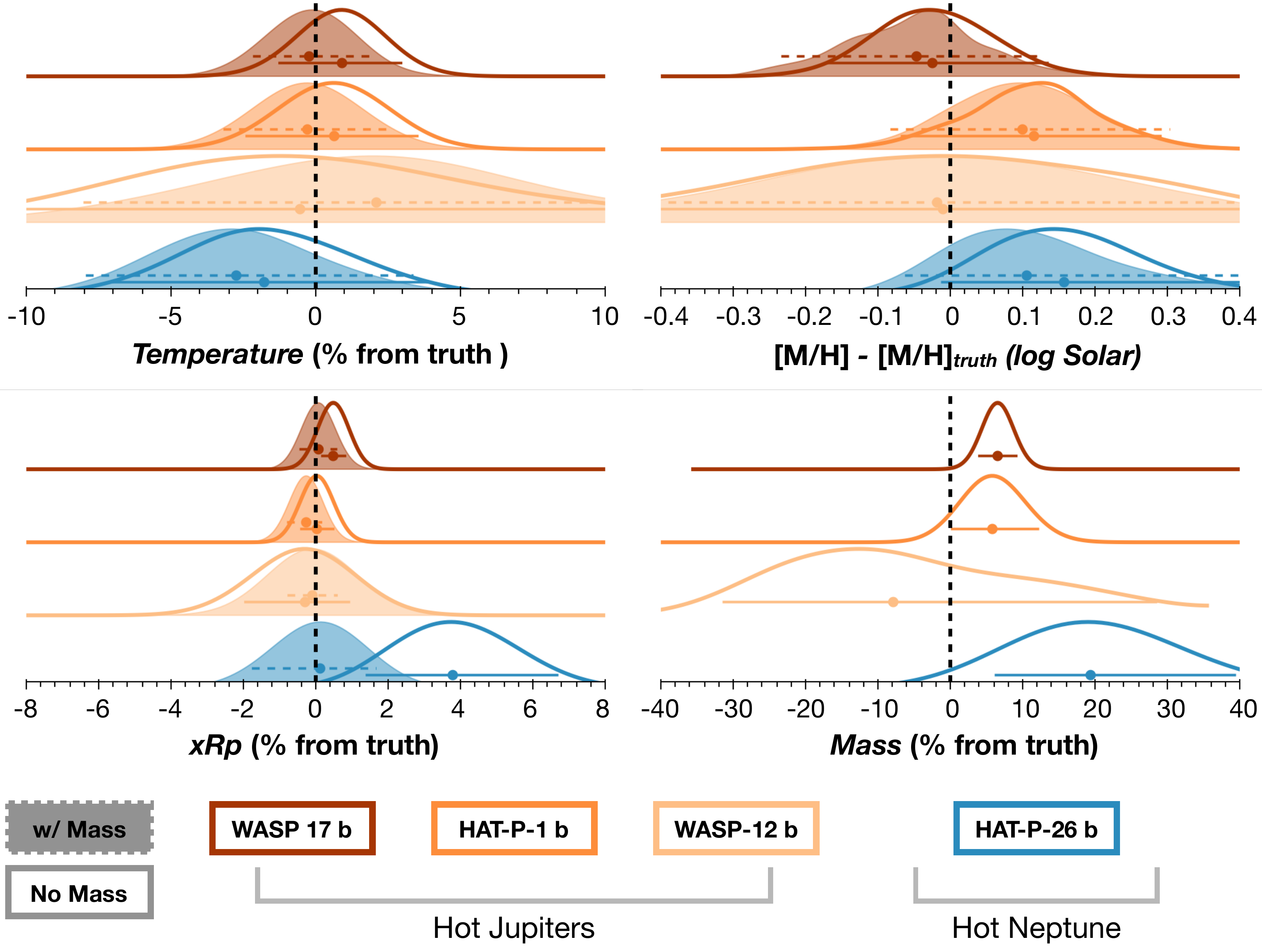}
\caption{Retrieved posterior distributions for the three hot Jupiters (WASP-17b, HAT-P-1b, and HD 189733 b) and the hot Neptune (HAT-P-26 b). The four retrieved parameters shown here (temperature, metallicity, reference radius, and mass) are a subset of the full state vector. All posteriors are shown relative to the true value. The shaded posteriors are for the retrieval cases where mass was fixed to the known value and the lines are when mass was a free parameter. \textbf{Main Point:} For the hot Jupiters, we find that the constraint on the atmospheric parameters is not largely dependent on whether or not mass is included as a free parameter.}
\label{fig:HJ} 
\end{figure*}

\section{Results}\label{sec:results}
For each case we start by retrieving atmospheric properties with a known mass measurement (i.e. disregarding mass as a free parameter entirely). Then, we include mass as a free parameter with a uniform prior of $\pm$20\%, $\pm$50\%, and $\pm$100\% the true mass value. Finally, we test the case of ``unknown'' mass, which we assume to be 0.1-30 M$_{Jup}$, 0.5-30 M$_{\oplus}$, and 0.1-30~M$_{\oplus}$ for the Jupiter-sized, Neptune-sized and terrestrial-sized planets, respectively. We compare each case by looking at the 2-$\sigma$ constraint interval and the mean of the retrieved posterior distribution, which we refer to as the precision and accuracy, respectively.

\subsection{WASP-17b, HAT-P-1b, WASP-12b, \& HAT-P-26b}
First we start with the three Hot Jupiters: WASP-17b, HAT-P-1b, and WASP-12b.  While all three systems were modeled with the same metallicity (1$\times$ Solar), the three had increasing levels of cloud cross sectional strength (log $\kappa_{cld}$ = -31, -30.5, and -29.5, respectively), and different temperatures and gravities. Figure~\ref{fig:HJ} shows a subset of the retrieved posterior distributions for temperature, metallicity, reference radius and mass for the case with a known mass (shaded), and the case with unknown mass as a free parameter (line). We do not show the intermediate cases  with priors of $\pm$20-100\% the true mass value because for these three planets, the retrieved precision and accuracy of the atmospheric parameters are statistically indistinguishable (within 1$\sigma$). This corroborates the analysis of \citet{dewit2013constraining} who was also able to retrieve atmospheric properties of hot Jupiters with unknown mass. 

Similar to \citet{welbanks2019degen} we do not find a significant degeneracy between reference radius and atmospheric parameters. We do find a degeneracy between reference radius and mass that leads to slightly overestimated mass constraints for two of the hot Jupiters ($+$6.6\% for WASP-17b and $+$5.8\% HAT-T-1b), and underestimation for the third hot Jupiter ($-$-12.5\% for WASP-12b). This degeneracy is not surprising given $g=GM/R(z)^2$ and any offset in reference radius leads to a slight offset in retrieved mass. 

Out of the three hot Jupiters, WASP-12b is the case with the highest degree of cloud coverage. This leads to spectral features that are highly muted, relative to WASP-17b and HAT-P-1b. Cloud-coverage is a well-documented effect that can sometimes impede precise retrievals of atmospheric parameters \citep{bet2016degen,bet2017degen,line2016clouds,macdonald2017hd209}. Therefore, WASP-12b exhibits wider retrieved posteriors for temperature, metallicity, reference radius, and mass, as shown in Figure~\ref{fig:HJ}, relative to the other two hot Jupiters.

Figure 2 also shows the case for hot Neptune HAT-P-26b. HAT-P-26b has a slightly enhanced metallicity (4$\times$Solar) compared with the hot Jupiter-cases (1$\times$Solar) and warm-Neptune-cases ($>$400$\times$ Solar). Therefore, it offers an opportunity to determine the effect of small changes in metallicity on the widths of the retrieved posterior distributions of mass, reference radius and the atmospheric parameters. While the precision of the retrieved metallicity and temperature distributions is not affected by an unknown mass, the accuracy of both are slightly degraded. Additionally, the accuracy and the precision of the retrieved reference radius are both highly impacted by the unknown mass. The +4\% overestimation of the reference radius leads to a +20\% overestimation of the mass. This degeneracy between mass and radius was seen as a minor effect in the hot Jupiter cases where metallicity was fixed at 1$\times$Solar. This foreshadows higher mean molecular weight as a potential inhibitor to constraining atmospheric properties, including mass, when mass is unknown. 

\begin{figure*}
\centering
\includegraphics[width=\textwidth]{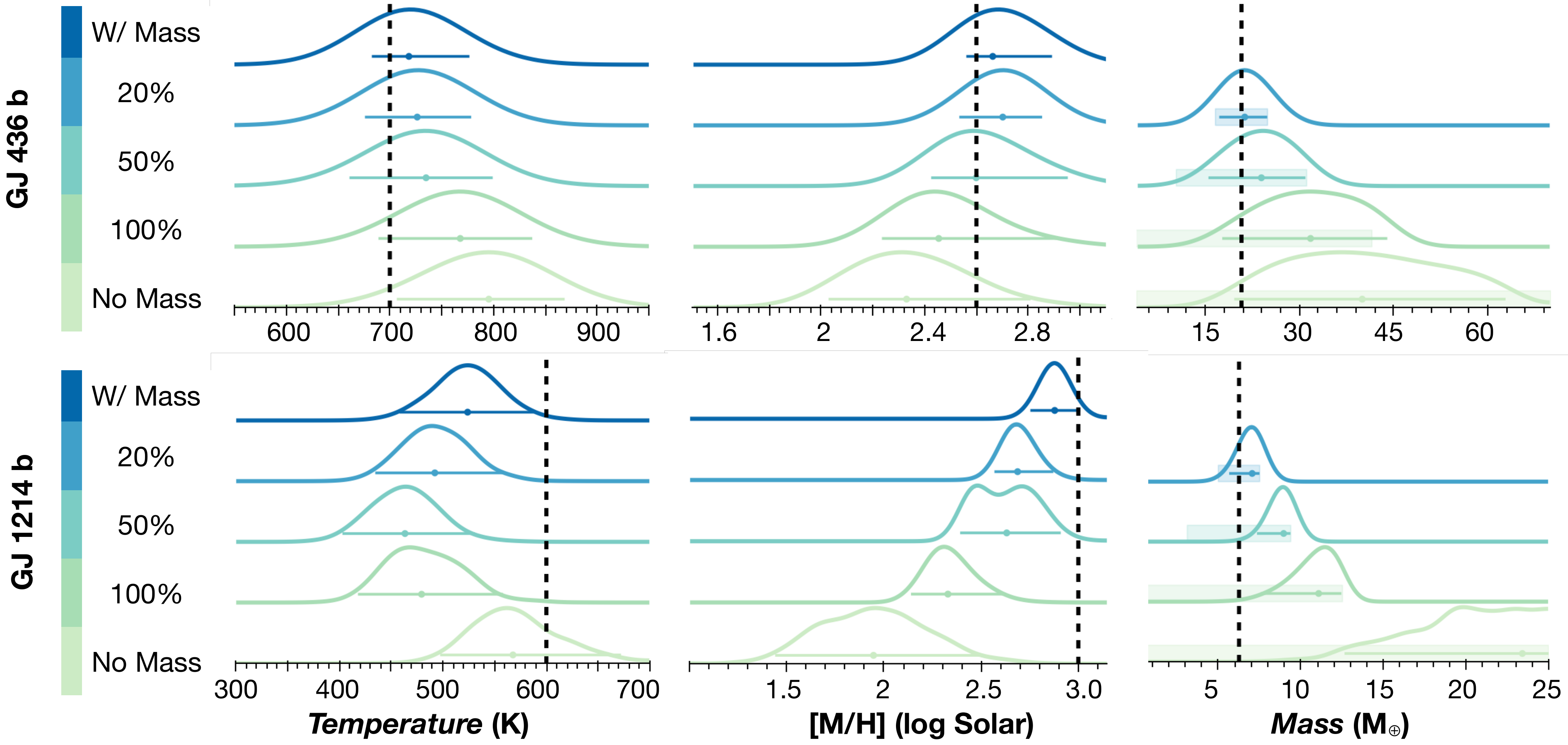}
\caption{Retrieved posterior distributions for GJ 436b and GJ 1214b. Dashed black lines indicate the true value. Horizontal colored lines show 2$\sigma$ constraint intervals. All posteriors are normalized to the same value. In the right most column, the shaded region shows the mass \textit{a priori} value used. The slight preference toward lower temperature and metallicity is a result of the random noise realization of the \textit{JWST} simulation. \textbf{Main Point:} The precision of the mass measurement dictates the accuracy and precision of the retrieved atmospheric properties. }
\label{fig:gj436} 
\end{figure*}

\begin{figure*}
\centering
\includegraphics[width=\textwidth]{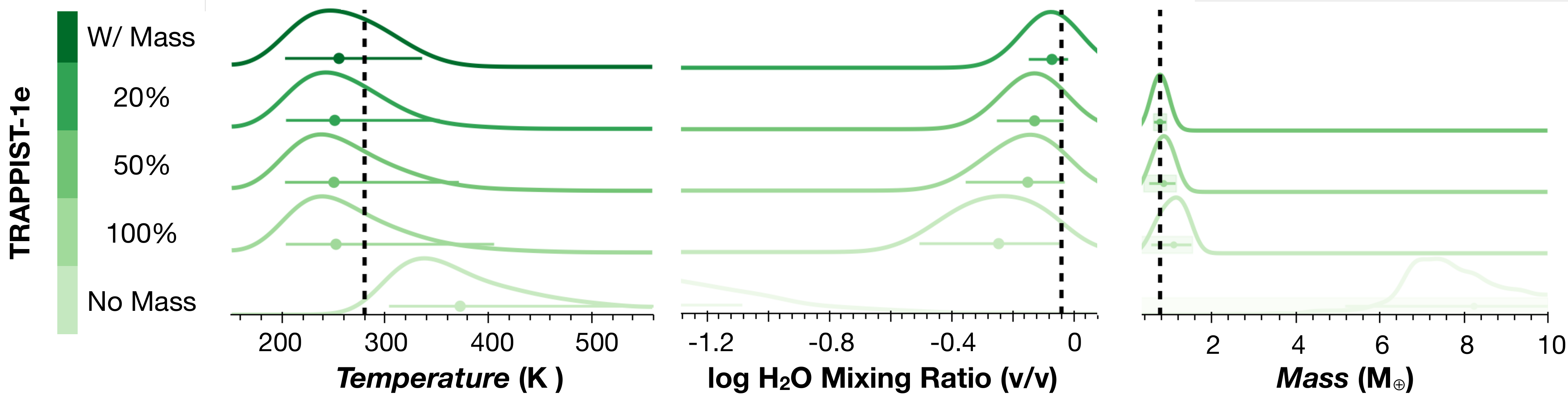}
\caption{Retrieved posterior distributions for TRAPPIST-1e. The schema of the figure is identical to Figure~\ref{fig:gj436}. \textbf{Main Point:}  We find that the precision of the mass measurement impacts the accuracy and precision of retrieved atmospheric properties most strongly for small planets like TRAPPIST-1e. }
\label{fig:trappy}
\end{figure*}

\subsection{GJ 436b \& GJ 1214b}
Next, we move to GJ 436b and GJ 1214b, a Neptune and a sub-Neptune, respectively. Figure~\ref{fig:gj436} shows the retrieved posterior probability distributions for all five cases (known mass, $\pm$20-100\% precision level masses, and unknown mass).

For GJ 436b, accurate metallicities can be retrieved in all cases (i.e. the true metallicity lies within 2$\sigma$ of the retrieved posterior probability distribution). However, the precision of the retrieved atmospheric parameters is dependent on how well the mass is known. If the mass is known to worse than $\pm$20\%, the precision on atmospheric parameters can degrade by up to a factor of $\sim$4. At the $\pm$20\% mass precision level, the metallicity constraint approaches that of when the mass is known (see Figure \ref{fig:gj436} top panel). The same effect is seen with temperature. 

For GJ 1214b, accurate and precise metallicities and temperatures cannot be retrieved in all cases. Here, if the mass is known to worse than $\pm$20\% the precision on atmospheric parameters can degrade by up to a factor of $\sim$5. We also find a degradation of accuracy of retrieved metallicity and temperature for cases where mass was known to worse than 20\% (clearly seen in Figure \ref{fig:gj436} bottom panel). The slight preference toward lower temperature and lower metallicity is a result of the random noise realization of the \textit{JWST} simulation. The bias is removed in a test case with 20 ppm error bars on each data point but zero Gaussian scatter added relative to the ground-truth forward model.

The right-most column of Figure~\ref{fig:gj436} shows the retrieved mass posterior distributions when mass was treated as a free parameter. The shaded region signifies the \textit{a prior} value of each case. In all cases the 2$\sigma$ upper bound was set by the \textit{a priori} value. In other words, it was not possible to constrain the mass or improve the upper bound of the mass from transmission spectroscopy. In a few cases, the lower bound of the 2$\sigma$ constraint could be slightly improved relative to the \textit{a priori} value. In these cases very low gravities could be excluded because the weak spectral features favored small scale heights, which corresponds to high gravities (H=$kT/\mu g$). 

One additional way that mass constraint could be improved, is by increasing the SNR of the spectrum. In order to test this we double the number of transit observations for GJ 436 and GJ 1214 to 4 and 6 per observing mode, respectively. We find that this improvement in SNR does not lead to significant changes in the precision or accuracy of the derived mass posteriors. Therefore we find that for these cases, mass constraints cannot be greatly improved from transmission spectroscopy alone.

\subsection{TRAPPIST-1e}
Lastly, we examine the results of the Earth-sized planet, TRAPPIST-1e, shown in Figure~\ref{fig:trappy}. Here we consider the best case scenario of a two gas mixture (H$_2$ and H$_2$O) in order to place limits on the impact of mass precision on the retrieved posterior distributions. 

The behavior of TRAPPIST-1e matches that of GJ 1214b. A degradation in the precision of mass, directly impacts the quality of the retrieved atmospheric parameters. The precision on the retrieved water abundance constraint is slightly impacted by the mass precision when it was known to $\pm$20\% level. If the mass is known to worse than $\pm$20\% the precision on atmospheric parameters can degrade by up to a factor of $\sim$23. Without a mass, only weak inferences can be made on the water abundance relative to the \textit{a priori} value because too many degenerate solutions exist. With regard to retrieving the masses directly, we find that the mass constraints cannot be greatly improved from transmission spectroscopy alone.

The overall magnitude of the H$_2$O and temperature constraint, shown in Figure~\ref{fig:trappy}, is optimistic compared with previous retrieval analysis of terrestrial atmospheres \citep{barstow2016hab, benneke2012atmospheric,kt2018detect} but comparable to other analyses of H$_2$O dominated atmospheres \citep[e.g.][]{Greene2016characterizing}. The difference comes from the low mean molecular weight of H$_2$O, as compared to a more Earth-like or a Venus-like atmosphere, that works to increase the scale height and strengthen the molecular features. 

\section{Discussion \& Conclusion}\label{sec:discon}
We have conducted a case study of seven planets to determine the level of planet mass precision required for robust atmospheric characterization. We have limited the analysis to focus on transmission spectroscopy with \textit{JWST} using NIRISS SOSS and NIRSpec G395H, which spans roughly 1-5~$\mu$m. Our sample consisted of three hot Jupiters, two warm Neptunes, one warm sub-Neptune, and one terrestrial-sized planet.  Simulations varied with regard to metallicity, cloud coverage, temperature, and gravity. For each planet, we conducted retrievals with varying levels of mass precision. In the control case, mass was not treated as a free parameter (i.e. known mass). In other cases, mass was allowed to vary in the retrieval by $\pm$20\%, $\pm$50\%, and $\pm$100\% relative to the true value. Lastly, mass was assumed to be unknown, with large uniform priors. For each retrieval, we determined how well atmospheric parameters were constrained in terms of: 1) whether the mean of the retrieved posterior probability distribution was centered on the true value (which we refer to as accuracy), and 2) the magnitude of the 2$\sigma$ constraint interval of each retrieved parameter (which we refer to as precision). Our conclusions are as follows: 
\begin{enumerate}
    \item \textbf{In all hot Jupiter cases} (which are 1$\times$~Solar metallicity), the retrieved precision and accuracy of the atmospheric parameters did not depend on whether or not mass was known. However, there was a slight degeneracy between retrieved reference radius and mass, which in some cases led to a $\sim$10\% overestimate of the true mass. Additionally, in the case of WASP-12b, which had the highest degree of cloud coverage, mass could only be retrieved with a precision of $\pm$30\% the true value. Therefore, even though atmospheric properties of (low metallicity) hot Jupiters can be determined, we should not expect to get precise or accurate mass measurements of hot Jupiters from transmission spectroscopy alone.
    \item \textbf{In the case of the hot Neptune}, HAT-P-26b, which is slightly enhanced in metals relative to solar (here, 4$\times$ Solar), the precision and accuracy of the retrieved metallicity and temperature were weakly dependent on whether or not mass was known \textit{a priori}. However, there was a degeneracy between reference radius and mass, which inhibited accurate retrievals of the planet's mass. 
    \item \textbf{In the case of the warm Neptunes and the Earth-sized planet}, the precision of the mass measurement dictated the precision with which atmospheric properties could be retrieved. At the $\pm$50\% mass precision level, accurate but not precise constraints on atmospheric properties could be attained. It was only at the $\pm$20\% mass precision level that the accuracy and precision of the retrieved atmospheric properties approached that of the control case where mass was known. These planets are also the planets that are most likely to lack mass measurements or to have high uncertainties in their measured masses.
\end{enumerate}

\subsection{Suggested Best Practices}
Although our analysis spanned several different planet types, it was certainly not exhaustive in terms of the full diversity of planets that have been discovered. Nevertheless we do not expect our results to change greatly for different sets of atmospheric assumptions and conclude this analysis by providing general community recommendations. We recommend caution in conducting any atmospheric characterization for planets lacking measured masses. We recommend a $\pm$50\% mass precision level for any initial atmospheric characterization with \textit{JWST} or \textit{Hubble}. Going into an atmospheric characterization investigation without any knowledge of a planet's mass limits our ability to place appropriate priors on the atmospheric properties.  For example, a low-density planet would be expected to have a very different atmosphere from a high density rocky planet of the same size.  Without prior knowledge of a planet's mass, it is therefore difficult to appropriately plan observations (in terms of e.g. making signal-to-noise predictions), on top of limiting our ability to correctly interpret the observations once they are obtained.

Lastly, although accurate atmospheric parameters can be retrieved at the $\pm$50\% level, the width of the posterior is impacted by the mass uncertainties. Therefore, for any detailed atmospheric characterization, we recommend masses be improved to $\pm$20\% so that the atmospheric constraints are only limited by the spectroscopic data quality itself. 

\acknowledgments
N.E.B acknowledges support from the University of California President’s Postdoctoral Fellowship Program. M.R.L. acknowledges support from the NASA Exoplanet Research Program award NNX17AB56G. The work of E.M.-R.K. was supported by the National Science Foundation under Grant No. 1654295 and by the Research Corporation for Science Advancement through their Cottrell Scholar program.

\software{numba \citep{numba}, bokeh \citep{bokeh}, NumPy \citep{walt2011numpy}, IPython \citep{perez2007ipython}, Jupyter, \citep{kluyver2016jupyter},PySynphot \citep{pysynphot2013}, PandExo \citep{pandexo}, dynesty \citep{dynesty}}




\begin{thebibliography}{}
\expandafter\ifx\csname natexlab\endcsname\relax\def\natexlab#1{#1}\fi
\providecommand{\url}[1]{\href{#1}{#1}}
\providecommand{\dodoi}[1]{doi:~\href{http://doi.org/#1}{\nolinkurl{#1}}}
\providecommand{\doeprint}[1]{\href{http://ascl.net/#1}{\nolinkurl{http://ascl.net/#1}}}
\providecommand{\doarXiv}[1]{\href{https://arxiv.org/abs/#1}{\nolinkurl{https://arxiv.org/abs/#1}}}

\bibitem[{{Amundsen} {et~al.}(2017){Amundsen}, {Tremblin}, {Manners},
  {Baraffe}, \& {Mayne}}]{amundsen2017cork}
{Amundsen}, D.~S., {Tremblin}, P., {Manners}, J., {Baraffe}, I., \& {Mayne},
  N.~J. 2017, \aap, 598, A97, \dodoi{10.1051/0004-6361/201629322}

\bibitem[{{Barstow} \& {Irwin}(2016)}]{barstow2016hab}
{Barstow}, J.~K., \& {Irwin}, P.~G.~J. 2016, \mnras, 461, L92,
  \dodoi{10.1093/mnrasl/slw109}

\bibitem[{Batalha {et~al.}(2019)Batalha, Stevenson, Fraine, Zhou, eas342, \&
  Cubillos}]{pandexo}
Batalha, N., Stevenson, K., Fraine, J., {et~al.} 2019, natashabatalha/PandExo:
  Pandeia/PandExo V1.4, \dodoi{10.5281/zenodo.3378022}.
\newblock \url{https://doi.org/10.5281/zenodo.3378022}

\bibitem[{{Batalha} {et~al.}(2017{\natexlab{a}}){Batalha}, {Kempton}, \&
  {Mbarek}}]{batalha2017challenges}
{Batalha}, N.~E., {Kempton}, E.~M.-R., \& {Mbarek}, R. 2017{\natexlab{a}},
  \apjl, 836, L5, \dodoi{10.3847/2041-8213/aa5c7d}

\bibitem[{{Batalha} \& {Line}(2017)}]{batalha2017ic}
{Batalha}, N.~E., \& {Line}, M.~R. 2017, \aj, 153, 151,
  \dodoi{10.3847/1538-3881/aa5faa}

\bibitem[{{Batalha} {et~al.}(2017{\natexlab{b}}){Batalha}, {Mandell},
  {Pontoppidan}, {Stevenson}, {Lewis}, {Kalirai}, {Earl}, {Greene}, {Albert},
  \& {Nielsen}}]{batalha2017pandexo}
{Batalha}, N.~E., {Mandell}, A., {Pontoppidan}, K., {et~al.}
  2017{\natexlab{b}}, \pasp, 129, 064501, \dodoi{10.1088/1538-3873/aa65b0}

\bibitem[{Benneke \& Seager(2012)}]{benneke2012atmospheric}
Benneke, B., \& Seager, S. 2012, The Astrophysical Journal, 753, 100

\bibitem[{{Benneke} \& {Seager}(2013)}]{benneke2013how}
{Benneke}, B., \& {Seager}, S. 2013, \apj, 778, 153,
  \dodoi{10.1088/0004-637X/778/2/153}

\bibitem[{{B{\'e}tr{\'e}mieux}(2016)}]{bet2016degen}
{B{\'e}tr{\'e}mieux}, Y. 2016, \mnras, 456, 4051, \dodoi{10.1093/mnras/stv2955}

\bibitem[{{B{\'e}tr{\'e}mieux} \& {Swain}(2017)}]{bet2017degen}
{B{\'e}tr{\'e}mieux}, Y., \& {Swain}, M.~R. 2017, \mnras, 467, 2834,
  \dodoi{10.1093/mnras/stx257}

\bibitem[{{Bokeh Development Team}(2014)}]{bokeh}
{Bokeh Development Team}. 2014, Bokeh: Python library for interactive
  visualization

\bibitem[{{de Wit} \& {Seager}(2013)}]{dewit2013constraining}
{de Wit}, J., \& {Seager}, S. 2013, Science, 342, 1473,
  \dodoi{10.1126/science.1245450}

\bibitem[{{de Wit} {et~al.}(2018){de Wit}, {Wakeford}, {Lewis}, {Delrez},
  {Gillon}, {Selsis}, {Leconte}, {Demory}, {Bolmont}, {Bourrier}, {Burgasser},
  {Grimm}, {Jehin}, {Lederer}, {Owen}, {Stamenkovi{\'c}}, \&
  {Triaud}}]{dewit2018trappy}
{de Wit}, J., {Wakeford}, H.~R., {Lewis}, N.~K., {et~al.} 2018, Nature
  Astronomy, 2, 214, \dodoi{10.1038/s41550-017-0374-z}

\bibitem[{Des~Etangs {et~al.}(2008)Des~Etangs, Pont, Vidal-Madjar, \&
  Sing}]{des2008rayleigh}
Des~Etangs, A.~L., Pont, F., Vidal-Madjar, A., \& Sing, D. 2008, Astronomy
  \&amp; Astrophysics, 481, L83

\bibitem[{{Fisher} \& {Heng}(2018)}]{fisher201838planets}
{Fisher}, C., \& {Heng}, K. 2018, \mnras, 481, 4698,
  \dodoi{10.1093/mnras/sty2550}

\bibitem[{{Freedman} {et~al.}(2014){Freedman}, {Lustig-Yaeger}, {Fortney},
  {Lupu}, {Marley}, \& {Lodders}}]{freedman2014opacities}
{Freedman}, R.~S., {Lustig-Yaeger}, J., {Fortney}, J.~J., {et~al.} 2014, \apjs,
  214, 25, \dodoi{10.1088/0067-0049/214/2/25}

\bibitem[{{Freedman} {et~al.}(2008){Freedman}, {Marley}, \&
  {Lodders}}]{freedman2008opacities}
{Freedman}, R.~S., {Marley}, M.~S., \& {Lodders}, K. 2008, \apjs, 174, 504,
  \dodoi{10.1086/521793}

\bibitem[{Greene {et~al.}(2016)Greene, Line, Montero, Fortney, Lustig-Yaeger,
  \& Luther}]{Greene2016characterizing}
Greene, T.~P., Line, M.~R., Montero, C., {et~al.} 2016, The Astrophysical
  Journal, 817, 17

\bibitem[{{Guillot}(2010)}]{guillot2010pt}
{Guillot}, T. 2010, \aap, 520, A27, \dodoi{10.1051/0004-6361/200913396}

\bibitem[{{Heng} \& {Kitzmann}(2017)}]{heng2017theory}
{Heng}, K., \& {Kitzmann}, D. 2017, \mnras, 470, 2972,
  \dodoi{10.1093/mnras/stx1453}

\bibitem[{Kluyver {et~al.}(2016)Kluyver, Ragan-Kelley, P{\'e}rez, Granger,
  Bussonnier, Frederic, Kelley, Hamrick, Grout, Corlay,
  {et~al.}}]{kluyver2016jupyter}
Kluyver, T., Ragan-Kelley, B., P{\'e}rez, F., {et~al.} 2016, in ELPUB, 87--90

\bibitem[{{Knutson} {et~al.}(2014){Knutson}, {Benneke}, {Deming}, \&
  {Homeier}}]{knutson2014gj436}
{Knutson}, H.~A., {Benneke}, B., {Deming}, D., \& {Homeier}, D. 2014, \nat,
  505, 66, \dodoi{10.1038/nature12887}

\bibitem[{{Kreidberg} {et~al.}(2014){Kreidberg}, {Bean}, {D{\'e}sert},
  {Benneke}, {Deming}, {Stevenson}, {Seager}, {Berta-Thompson}, {Seifahrt}, \&
  {Homeier}}]{kreidberg2014gj1214}
{Kreidberg}, L., {Bean}, J.~L., {D{\'e}sert}, J.-M., {et~al.} 2014, \nat, 505,
  69, \dodoi{10.1038/nature12888}

\bibitem[{{Kreidberg} {et~al.}(2015){Kreidberg}, {Line}, {Bean}, {Stevenson},
  {D{\'e}sert}, {Madhusudhan}, {Fortney}, {Barstow}, {Henry}, {Williamson}, \&
  {Showman}}]{kreidberg2015wasp12}
{Kreidberg}, L., {Line}, M.~R., {Bean}, J.~L., {et~al.} 2015, \apj, 814, 66,
  \dodoi{10.1088/0004-637X/814/1/66}

\bibitem[{{Krissansen-Totton} {et~al.}(2018){Krissansen-Totton}, {Garland},
  {Irwin}, \& {Catling}}]{kt2018detect}
{Krissansen-Totton}, J., {Garland}, R., {Irwin}, P., \& {Catling}, D.~C. 2018,
  \aj, 156, 114, \dodoi{10.3847/1538-3881/aad564}

\bibitem[{Lam {et~al.}(2015)Lam, Pitrou, \& Seibert}]{numba}
Lam, S.~K., Pitrou, A., \& Seibert, S. 2015, in Proceedings of the Second
  Workshop on the LLVM Compiler Infrastructure in HPC, LLVM '15 (New York, NY,
  USA: ACM), 7:1--7:6.
\newblock \url{http://doi.acm.org/10.1145/2833157.2833162}

\bibitem[{Line {et~al.}(2014)Line, Knutson, Wolf, \& Yung}]{line2014systematic}
Line, M.~R., Knutson, H., Wolf, A.~S., \& Yung, Y.~L. 2014, The Astrophysical
  Journal, 783, 70

\bibitem[{{Line} \& {Parmentier}(2016)}]{line2016clouds}
{Line}, M.~R., \& {Parmentier}, V. 2016, \apj, 820, 78,
  \dodoi{10.3847/0004-637X/820/1/78}

\bibitem[{{Line} {et~al.}(2012){Line}, {Zhang}, {Vasisht}, {Natraj}, {Chen}, \&
  {Yung}}]{line2012info}
{Line}, M.~R., {Zhang}, X., {Vasisht}, G., {et~al.} 2012, \apj, 749, 93,
  \dodoi{10.1088/0004-637X/749/1/93}

\bibitem[{Line {et~al.}(2013)Line, Wolf, Zhang, Knutson, Kammer, Ellison,
  Deroo, Crisp, \& Yung}]{line2013systematic}
Line, M.~R., Wolf, A.~S., Zhang, X., {et~al.} 2013, The Astrophysical Journal,
  775, 137

\bibitem[{{Lupu} {et~al.}(2014){Lupu}, {Zahnle}, {Marley}, {Schaefer},
  {Fegley}, {Morley}, {Cahoy}, {Freedman}, \& {Fortney}}]{lupu2014earth}
{Lupu}, R.~E., {Zahnle}, K., {Marley}, M.~S., {et~al.} 2014, \apj, 784, 27,
  \dodoi{10.1088/0004-637X/784/1/27}

\bibitem[{{MacDonald} \& {Madhusudhan}(2017)}]{macdonald2017hd209}
{MacDonald}, R.~J., \& {Madhusudhan}, N. 2017, \mnras, 469, 1979,
  \dodoi{10.1093/mnras/stx804}

\bibitem[{{Mandell} {et~al.}(2013){Mandell}, {Haynes}, {Sinukoff},
  {Madhusudhan}, {Burrows}, \& {Deming}}]{mandell2013wasp17}
{Mandell}, A.~M., {Haynes}, K., {Sinukoff}, E., {et~al.} 2013, \apj, 779, 128,
  \dodoi{10.1088/0004-637X/779/2/128}

\bibitem[{McBride \& Gordon(1996)}]{mcbride1996computer}
McBride, B.~J., \& Gordon, S. 1996

\bibitem[{P{\'e}rez \& Granger(2007)}]{perez2007ipython}
P{\'e}rez, F., \& Granger, B.~E. 2007, Computing in Science \& Engineering, 9

\bibitem[{{Rocchetto} {et~al.}(2016){Rocchetto}, {Waldmann}, {Venot}, {Lagage},
  \& {Tinetti}}]{rocchetto2016bias}
{Rocchetto}, M., {Waldmann}, I.~P., {Venot}, O., {Lagage}, P.-O., \& {Tinetti},
  G. 2016, \apj, 833, 120, \dodoi{10.3847/1538-4357/833/1/120}

\bibitem[{{Schlawin} {et~al.}(2018){Schlawin}, {Greene}, {Line}, {Fortney}, \&
  {Rieke}}]{schlawin2018clear}
{Schlawin}, E., {Greene}, T.~P., {Line}, M., {Fortney}, J.~J., \& {Rieke}, M.
  2018, \aj, 156, 40, \dodoi{10.3847/1538-3881/aac774}

\bibitem[{{Sing} {et~al.}(2016){Sing}, {Fortney}, {Nikolov}, {Wakeford},
  {Kataria}, {Evans}, {Aigrain}, {Ballester}, {Burrows}, {Deming},
  {D{\'e}sert}, {Gibson}, {Henry}, {Huitson}, {Knutson}, {Lecavelier Des
  Etangs}, {Pont}, {Showman}, {Vidal-Madjar}, {Williamson}, \&
  {Wilson}}]{sing2016hjs}
{Sing}, D.~K., {Fortney}, J.~J., {Nikolov}, N., {et~al.} 2016, \nat, 529, 59,
  \dodoi{10.1038/nature16068}

\bibitem[{{Speagle}(2019)}]{dynesty}
{Speagle}, J.~S. 2019, arXiv e-prints.
\newblock \doarXiv{1904.02180}

\bibitem[{{STScI Development Team}(2013)}]{pysynphot2013}
{STScI Development Team}. 2013, {pysynphot: Synthetic photometry software
  package}, Astrophysics Source Code Library.
\newblock \doeprint{1303.023}

\bibitem[{{Wakeford} {et~al.}(2013){Wakeford}, {Sing}, {Deming}, {Gibson},
  {Fortney}, {Burrows}, {Ballester}, {Nikolov}, {Aigrain}, {Henry}, {Knutson},
  {Lecavelier des Etangs}, {Pont}, {Showman}, {Vidal-Madjar}, \&
  {Zahnle}}]{wakeford2013hatp1}
{Wakeford}, H.~R., {Sing}, D.~K., {Deming}, D., {et~al.} 2013, \mnras, 435,
  3481, \dodoi{10.1093/mnras/stt1536}

\bibitem[{{Wakeford} {et~al.}(2017){Wakeford}, {Sing}, {Kataria}, {Deming},
  {Nikolov}, {Lopez}, {Tremblin}, {Amundsen}, {Lewis}, {Mandell}, {Fortney},
  {Knutson}, {Benneke}, \& {Evans}}]{wakeford2017hatp26}
{Wakeford}, H.~R., {Sing}, D.~K., {Kataria}, T., {et~al.} 2017, Science, 356,
  628, \dodoi{10.1126/science.aah4668}

\bibitem[{Walt {et~al.}(2011)Walt, Colbert, \& Varoquaux}]{walt2011numpy}
Walt, S. v.~d., Colbert, S.~C., \& Varoquaux, G. 2011, Computing in Science \&
  Engineering, 13, 22

\bibitem[{{Welbanks} \& {Madhusudhan}(2019)}]{welbanks2019degen}
{Welbanks}, L., \& {Madhusudhan}, N. 2019, \aj, 157, 206,
  \dodoi{10.3847/1538-3881/ab14de}

\end{thebibliography}
\end{document}